\documentstyle[jkas]{article}

\beginpage{1}
\endpage{4}
\year{2011}\volume{??}\month{April}\issueno{2}

\runningauthor {J.-E. LEE ET AL.}
\runningtitle{OPTICAL SPECTRA OF HBC 722}

\month{April} \year{2011} \volume{44} \issueno{2}
\beginpage{1}\endpage{4}
\date{Received March 21, 2011; Revised April 01, 2011; Accepted April 01, 2011}

\begin{document}
\title{HIGH RESOLUTION OPTICAL SPECTRA OF HBC 722 AFTER OUTBURST}
\author{Jeong-Eun Lee$^{1}$, Wonseok Kang$^{1}$, Sang-Gak Lee$^{2}$, Hyun-Il Sung$^{3}$, Byeong-Cheol Lee$^{3}$, Hwankyung Sung$^{4}$, Joel D. Green$^{5}$, and Young-Beom Jeon$^{3}$}
\address{$^1$ Department of Astronomy and Space Science, Kyung Hee University,
  Yongin-si, Gyeonggi-do 446-701, Korea\\
 {\it E-mail : jeongeun.lee@khu.ac.kr, wskang@astro.snu.ac.kr}}
\address{$^2$ Department of Physics and Astronomy, Seoul National University,
Seoul 151-742, Korea\\
 {\it E-mail : sanggak@snu.ac.kr}}
\address{$^3$ Korea Astronomy and Space Science Institute,
61-1 Whaam-dong, Yuseong-Gu, Daejeon 305-348, Korea\\
  {\it E-mail : hisung@kasi.re.kr, bclee@kasi.re.kr, ybjeon@kasi.re.kr}}
 \address{$^4$ Department of Astronomy and Space Science, Sejong University, Kunja-dong 98, Kwangjin-gu, Seoul 143-747, Korea\\
  {\it E-mail : sungh@sejong.ac.kr}}
 \address{$^5$ Department of Astronomy, University of Texas at Austin, 1 University Station C1400, Austin, TX 78712, USA\\
  {\it E-mail : joel@astro.as.utexas.edu}}

\offprints{J.-E. Lee}
\abstract{
We report the results of our high resolution optical spectroscopic monitoring campaign ($\lambda$ $=$ 3800 -- 8800 \AA, R $=$ 30000 -- 45000) of the new FU Orionis-type object HBC 722.  We observed HBC 722 with the BOES 1.8-m telescope between November 26 and December 29, 2010, and FU Orionis itself on January 26, 2011.  We detect a number of previously unreported high-resolution K I and Ca II lines beyond 7500 \AA.  We resolve the H$\alpha$ and Ca II line profiles into three velocity components, which we attribute to both disk and outflow. The increased accretion during outburst can heat the disk to produce the relatively narrow absorption feature and launch outflows appearing as high velocity blue and red-shifted broad features.}

\maketitle

\section{INTRODUCTION}

\begin{table*}[!!t]
\begin{center}
\centering
\caption{Observation log \label{tbl1} }
\begin{tabular}{lccccccc}
\hline \hline
   Target & Date & UT & Exposure time & Binning &  Fiber    & Spatial Resolution$^{\rm a}$ & Seeing  \\
          &      &    &   (sec)       &         &  ($\mu$m) &                              &         \\
\hline
   HBC 722 & 2010 Nov 26 & 09:44 & 3600 & 2$\times$2  & 300  &   4.3$\arcsec$  &   2.6$\arcsec$ \\
   HBC 722 & 2010 Dec 11 & 09:09 & 3600 & 1$\times$1  & 300  &   4.3$\arcsec$  &   3.3$\arcsec$ \\
   HBC 722 & 2010 Dec 23 & 09:16 & 3600 & 2$\times$2  & 200  &   2.9$\arcsec$  &   3.5$\arcsec$ \\
   HBC 722 & 2010 Dec 29 & 09:40 & 3600 & 2$\times$2  & 200  &   2.9$\arcsec$  &   3.5$\arcsec$ \\
   FU Orionis$^{\rm b}$  & 2011 Jan 26 & 13:01 & 3600 & 1$\times$1  & 200 &  2.9$\arcsec$ & 3.7$\arcsec$ \\
\hline
\end{tabular}
\end{center}
\begin{tabnote}
 \hskip18pt $^{\rm a}$ Spatial resolution corresponding to the fiber size. This value means the fiber-tip field-of-view.  \\
\end{tabnote}
\begin{tabnote}
 \hskip18pt $^{\rm b}$ The coordinates of FU Orionis in RA and DEC are respectively, (5h 45m 22.4s, $+09^\circ\ 04\arcmin\ 12.4\arcsec$, J2000). \\
\end{tabnote}
\end{table*}
\begin{table}[!t]
\begin{center}
\centering
\caption{Detected line list from high-resolution spectra \label{tbl2} }
\begin{tabular}{lccc}
\hline \hline
  Elem. & $\lambda$ & This & Miller \\
        &   (\AA)   & Work & \etal  \\
\hline
   Fe II     & 5018.43 & no  & yes \\
   Mg I      & 5167.32 & no  & yes \\
   Fe II     & 5169.03 & no  & yes \\
   Mg I      & 5172.63 & yes & yes \\
   Mg I      & 5183.60 & yes & yes \\
   Na I      & 5889.95 & yes & yes \\
   Na I      & 5895.92 & yes & yes \\
   Ba II     & 6141.71 & no  & yes \\
   Ba II     & 6496.90 & no  & yes \\
   Fe I      & 6393.60 & no  & yes \\
   Fe I      & 6592.91 & no  & yes \\
   Fe I      & 6677.99 & no  & yes \\
   H$\alpha$ & 6562.79 & yes & yes \\
   K I       & 7664.90 & yes & no  \\
   K I       & 7698.96 & yes & no  \\
   Ca II     & 8498.02 & yes & no  \\
   Ca II     & 8542.09 & yes & no  \\
   Ca II     & 8662.14 & yes & no \\
\hline
\end{tabular}
\end{center}
\end{table}

The standard star formation model predicts a constant accretion rate \citep{shu77,tsc84,shu87}.
However, recent studies based on surveys toward nearby low-mass star forming regions (Dunham \etal\ 2010, and references therein) suggest that the luminosities of young stellar objects are systematically low compared to the standard model.
In addition, the discovery of Very Low Luminosity Objects (i.e. VeLLOs; Young \etal\ 2004; Bourke \etal\ 2006) and their associated strong outflows \citep{andre99,lee10} raised questions about the steady accretion process. As a result, an alternate mechanism termed the episodic accretion process has been suggested, to account for these observational phenomena (Lee 2007, and references therein).
The episodic accretion process is characterized by two phases: burst and quiescent accretion.
FU Orionis-type objects (hereafter, FUors) have been proposed as prominent examples of burst accreting protostars, while VeLLOs have been proposed as objects in the quiescent phase of the episodic accretion process.

FUors are a class of low-mass pre-main sequence objects named after FU Orionis, which produced a 5 magnitude optical outburst in 1936 and has remained in its brightened state. As a consequence of eruptive accretion, these protostars exhibit large winds and outflows (Croswell, Hartmann, \& Avrett 1987), which are inferred from P Cygni profiles of H$\alpha$ and other lines.
The spectral characteristics of FUors are broad blueshifted emission lines, IR excess, and near-IR CO overtone features, consequences of the energetic burst of accretion-driven viscous heating of the disk.  Based on these characteristics, \citet{har96} and \citet{rei10} identified about dozen FUors, although in many cases the initial outburst had not been observed.   Very little pre-outburst data exists for FUors; few have been studied from the pre-burst phase to the burst phase, and only one (V1057 Cyg) has a pre-outburst optical spectrum.

HBC 722, also known as LkH$\alpha$ 188-G4 and PTF10qpf, was recently identified as a FUor by \citet{sem10} and \citet{mil11}.
Its brightness excursion is $\Delta$V$=$4.7 mag \citep{sem10}, reaching its maximum brightness in September 2010, slowly decreasing since then.
HBC 722 (RA $=$ 20h 58m 17.0s, Dec $=+43^\circ\ 53\arcmin\ 42.9\arcsec$, J2000) is located in an active star forming region of North American/Pelican Nebula, 520 pc away \citep{lau06}.
Pre-outburst, HBC 722 was identified as an emission line object of spectral type K7-M0 in the classical T Tauri phase, with an interstellar reddening of $A_V =$ 3.4 mag \citep{coh79}.
During the burst, the optical and NIR spectra of HBC 722 show consistency with a G-type giant/supergiant and M-type giant/supergiant, respectively \citep{mil11}.
The burst has not yet exhibited far-infrared feedback, detectable with the instruments on board the Herschel Space Observatory \citep{gre11}.  HBC 722 is the best-characterized FUor-like object pre-outburst, and it provides the first opportunity to profile the burst phase of accretion across all wavelengths, allowing us to model the process in detail.

Here we report the high-resolution optical spectra of HBC 722, observed about two months after reaching the maximum brightness of its outburst.

\begin{figure*}[!!t]
\epsscale{1.8}
\plotone{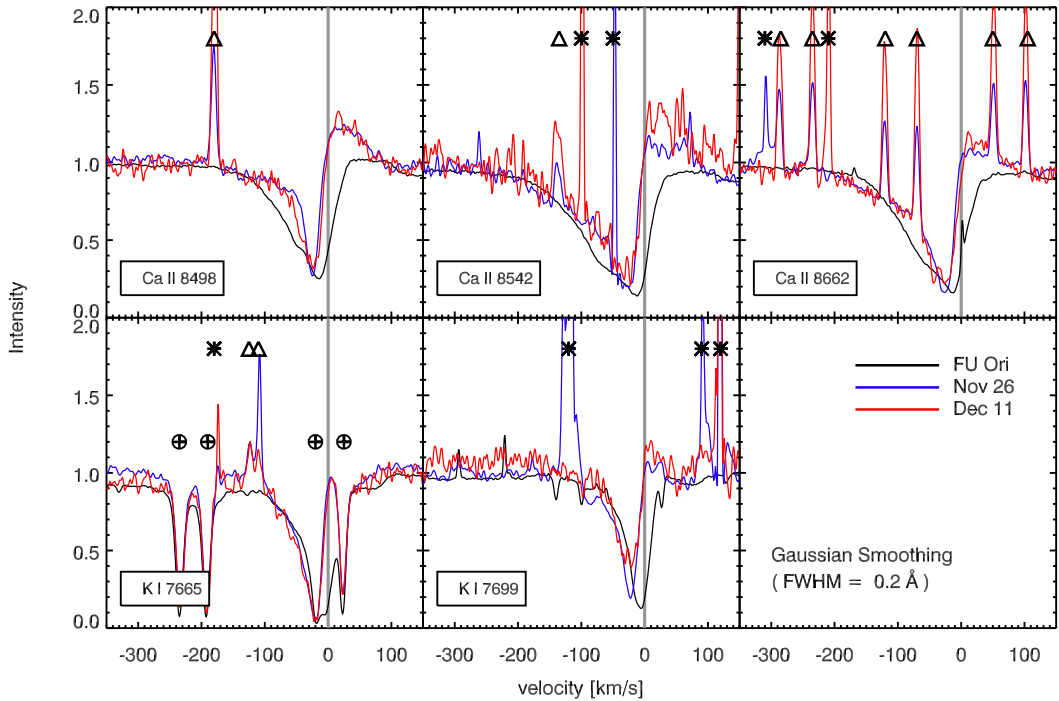}
\caption{
Ca II and K I lines observed in spectra taken on November 26 (blue) and December 11 (red), 2010.
All spectra are shown at the resolution smoothed to 0.2 \AA. The spectrum of FU Orionis (black) is also shown for comparison.
Line intensity is normalized to the continuum level. Telluric emission lines (triangles; Hanuschik 2003), emission lines caused
by hot pixels (asterisks), and telluric absorption (earth symbols) are marked.  \label{fig1}}
\end{figure*}


\begin{figure}[!!t]
\epsscale{1.0}
\plotone{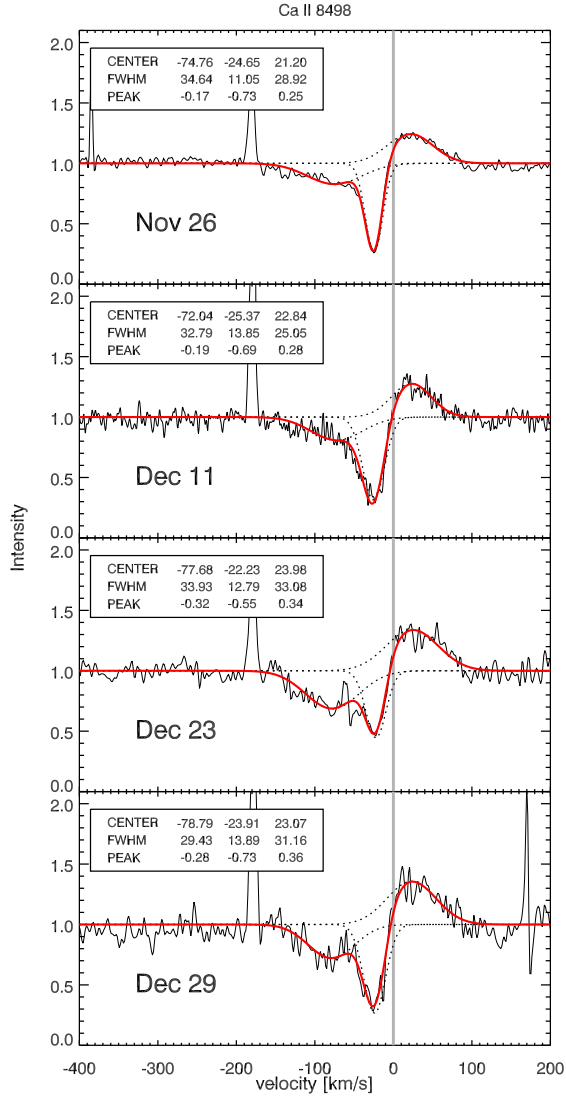}
\caption{The Ca II 8498 spectra fitted with three Gaussian profiles: two absorption components at $v_{center} \sim -75$ and $-25$ ${\rm km~s^{-1}} $ and an emission component around $v_{center} \sim +25$ ${\rm km~s^{-1}}$. The black dotted lines indicate fitted Gaussian profiles, and the red solid lines represent the whole combined profiles. Fit parameters including central velocity, peak intensity, and velocity FWHM are presented in the legend boxes. \label{fig2}}
\end{figure}

\begin{figure}[!!t]
\epsscale{1.0}
\plotone{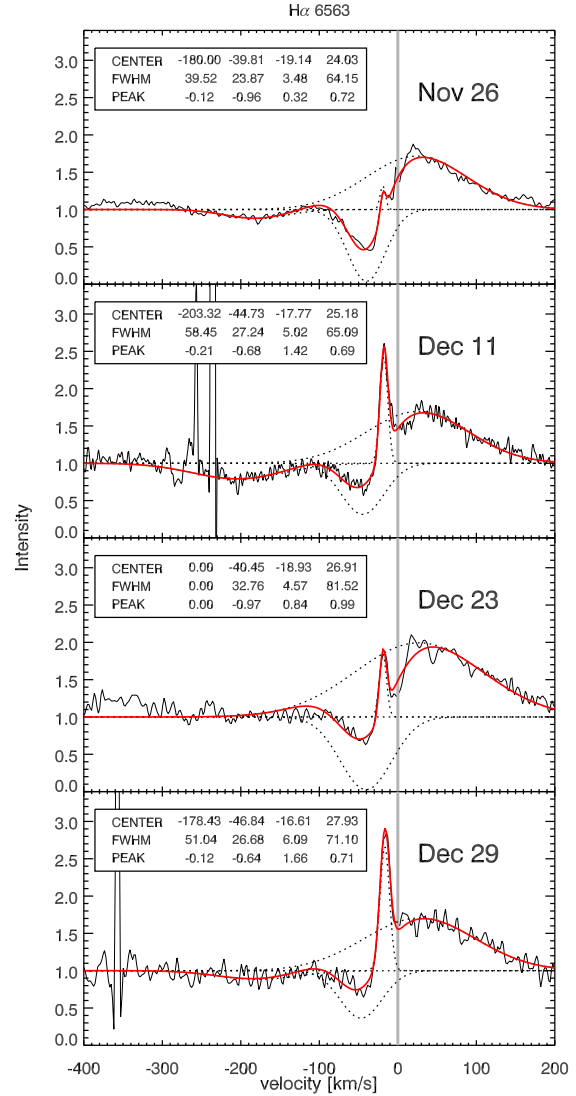}
\caption{The H$\alpha$ spectra fitted with four Gaussian profiles: two absorption components around $v_{center} \sim -200$ and $-45$ ${\rm km~s^{-1}} $, an emission component around $v_{center} \sim +25~{\rm km~s^{-1}}$, and a telluric emission line. Line shapes and legends are as in \figurename~\ref{fig2}. \label{fig3}}
\end{figure}

\begin{figure}[!!t]
\epsscale{1.0}
\plotone{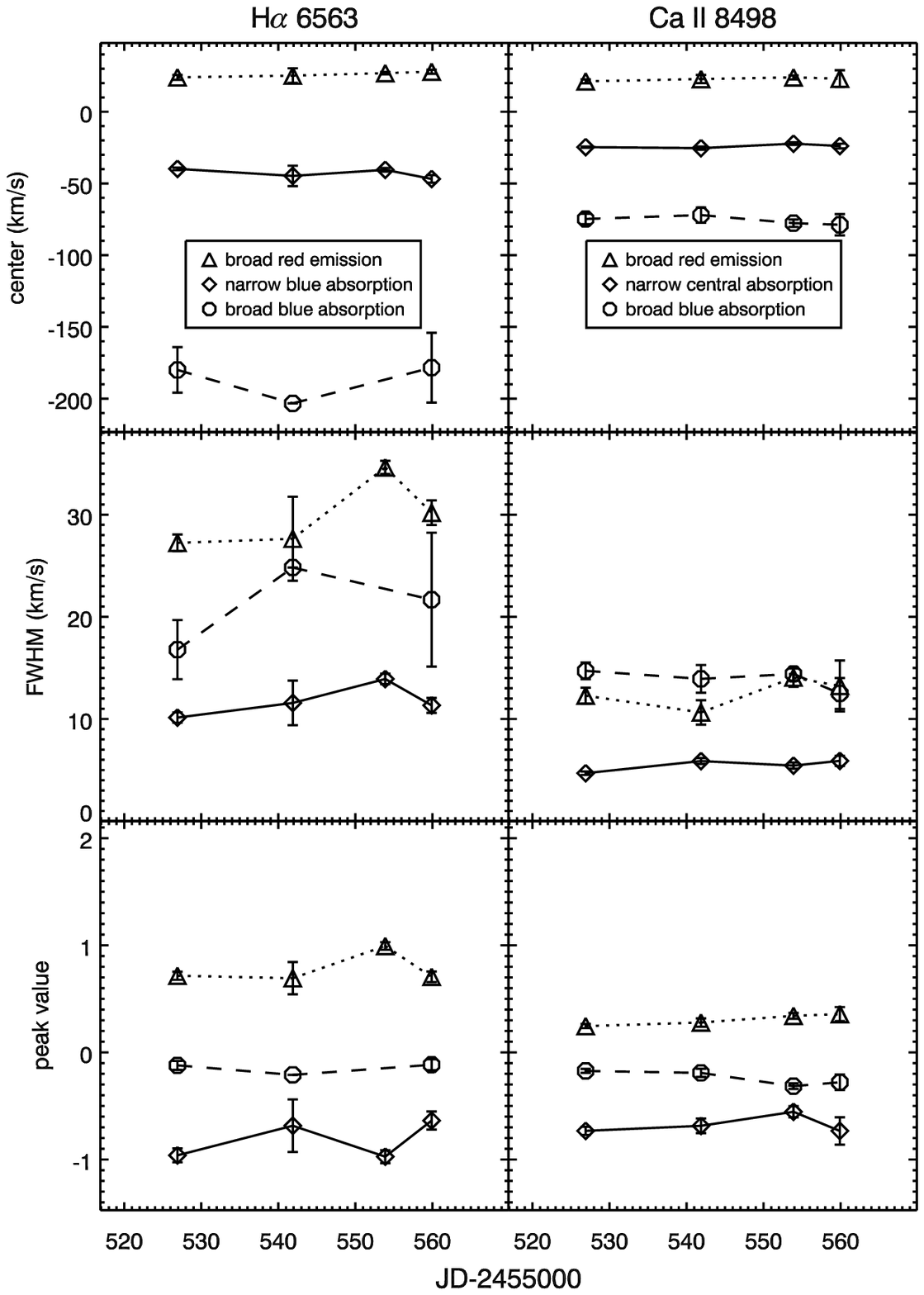}
\caption{Time variation of each fitted parameter for the
H$\alpha$ and Ca II 8498 lines. Different symbols indicate the different velocity components of the lines (identified in legend boxes).
Error bars represent formal 1$\sigma$ errors. \label{fig4}}
\end{figure}
\section{OBSERVATIONS}

We have carried out optical high-resolution spectroscopic observations of HBC 722 from November 26 till December 29, 2010 using the Bohyunsan Optical Echell Spectrograph (BOES; Kim \etal\ 2002, 2007) attached to the 1.8 m telescope in the Bohyunsan Optical Astronomy Observatory (BOAO) in Korea.
We have also observed FU Orionis itself for comparison on January 26, 2011.
All spectra were obtained with BOES using either the 200 (R $\sim$ 45,000) or the 300 $\mu$m (R $\sim$ 30,000) fiber.
The observed spectral regions cover the optical bands in the 3800 -- 8800 \AA\ range.
The typical signal-to-noise ratio at 6700 \AA\ is $\sim$ 15.
The observation log is listed in \tablename~\ref{tbl1}.

The observed spectra were reduced with the IRAF \texttt{echelle} package to produce the spectra for each order of the echelle spectrum.
The echelle aperture tracing was performed using the master flat image, a combination of all flat images. After aperture tracing, the flat, the comparison, and the object spectra were extracted from each image, with the same aperture reference as the master flat image. In the flatfielding process, the interference fringes and the pixel-to-pixel spectral variations were corrected.  Wavelength calibration was performed with the ThAr lamp spectrum, and the object spectra were normalized in each aperture using the \texttt{continuum} task.

\section{ANALYSIS}

First we identified lines from our BOES spectra of HBC 722, using the spectra presented in \cite{mil11} as a template.  At wavelengths greater than 5000 \AA, the BOES spectra have relatively high S/N ratio, even at wavelengths greater than 8000 \AA. However, lines located at wavelengths shorter than 5000 \AA\ were difficult to identify because of low S/N ratio in this region.
The line comparison between our spectra and those presented in \cite{mil11} is listed in \tablename~\ref{tbl2}.  \figurename~\ref{fig1} shows two K I lines near 7700 \AA\, and the Ca II triplet lines near 8500 \AA.

The spectra of FU Orionis are also plotted for comparison.
Although these lines were detected with low-resolution spectroscopy (Miller et al. 2011), the high-resolution spectra presented here show clear
blue-shifted absorption and red-shifted emission components; this P Cygni profile was previously reported only in H$\alpha$.

We selected two lines with strong P Cygni profiles, H$\alpha$ and Ca II 8498, to perform a least-$\chi^2$ fit with Gaussian profiles in order to examine the time-variations of central velocity, FWHM, and peak intensity for each component.
The fitting results are shown in \figurename~\ref{fig2} and \figurename~\ref{fig3}.

The Ca II 8498 lines can be decoupled into three components: a broad blue absorption feature, a broad red emission feature, and a relatively narrow central absorption feature.
The broad blue absorption and red emission features are most likely associated with outflows, while the relatively narrow absorption feature may be associated with the disk.
If the narrow absorption profile is really produced by the disk, we can determine the radial velocity of HBC 722 as the central velocity of the profile, which has a small variation between
$-22$ to $-25$ ${\rm km~s^{-1}} $. This idea is supported by the fact that the central velocities of two outflow components differ by about $\pm50$ ${\rm km~s^{-1}} $ from the central velocity of the narrow absorption profile, which is considered as the disk component.

The H$\alpha$ lines are also decoupled into three velocity components as seen in
\figurename~\ref{fig3}: a very broad but shallow blue-shifted absorption feature, a relatively narrow
blue absorption feature, and a very broad red emission feature.
The very narrow emission feature superimposed on the broad emission one is a telluric emission line (Hanuschik 2003),
which was included in the fitting of the Gaussian profiles.
The relatively narrow absorption feature might be associated with the disk component, as suggested from the Ca II line.
The central velocity of the very broad absorption feature is about $-200$ ${\rm km~s^{-1}} $, which is possibly the same fast outflow component as detected from the H$\gamma$ line in \cite{mil11}. However,
this very broad blue absorption feature was not detected on 23 December, 2010,
probably due to a very low S/N in this spectrum. The very broad blue absorption and red emission features must be also associated with outflows, indicative of different velocity components within the outflow. The broad red emission line is very bright compared to the blue absorption line; this large intensity difference suggests that the outflow spans a large area.

In \figurename~\ref{fig4}, we plot the time variation of each parameter of these fitted Gaussian profiles in
H$\alpha$ and Ca II 8498 lines. The variation is not significant in most components, given the low S/N ratio; the only exception is the FWHM of the broad red emission feature of H$\alpha$, which varied considerably during our observations.
In addition, the peak intensity of the broad red emission feature seems to correlate with the variation of its FWHM.
The variation in the outflowing material is typically associated with a varying mass accretion rate (e.g. Kurosawa \etal\ 2006). Therefore, in order to understand the accretion process post-outburst, we plan to monitor the variation of each velocity component  when HBC 722 becomes observable again, in April 2011.

\section{SUMMARY}

We present a time series of BOES high-resolution optical spectra of HBC 722, a newly reported FUor-like outburst during the summer of 2010. The high resolution spectra of two K I lines near 7700 \AA\ and the Ca II triplet lines near 8500 \AA\ were covered uniquely by our observations.
The P Cygni profiles were clearly detected in the H$\alpha$ and Ca II lines, but only marginally in the K I lines. Using Gaussian fits to the observed line profiles, we show that the Ca II 8498 and the H$\alpha$ 6563 lines trace the disk component as well as the outflow. The highest velocity component of the outflow ($\sim -200$ ${\rm km~s^{-1}} $) was detected in the H$\alpha$ line, and the broad red emission feature of the H$\alpha$ line varies the most. In the future, monitoring the variation of spectral features will help our understanding of the kinematic structures and of their time variations, as HBC 722 returns to its pre-outburst state.

\acknowledgments{
This research was supported by the Basic Science Research Program through the National Research Foundation of Korea (NRF) funded by the Ministry of Education, Science and Technology (No. 2010-0008704)}

\end{document}